\documentclass[aip,jcp,amsmath,amssymb,preprint]{revtex4-1}
\usepackage{graphicx}% Include figure files
\usepackage{dcolumn}% Align table columns on decimal point
\usepackage{bm}% bold math
\usepackage{mhchem}
\usepackage{tabularx}
\usepackage{gensymb}
\usepackage{threeparttable}
\usepackage{float}

\newcommand{\kj}{kJ mol$^{-1}$}
\newcommand{\dGhydr}{$\Delta G_\textrm{hydr}^\circ$}

\begin{document}
\graphicspath{{}}
\preprint{APS/123-QED}

\title{The Hydration Structure of Carbon Monoxide by Ab Initio Methods}

\author{Ernest Awoonor-Williams}
\author{Christopher N. Rowley}
\email{crowley@mun.ca}
\affiliation{Department of Chemistry, Memorial University of Newfoundland, St. John's, NL, Canada}%

\date{\today}% It is always \today, today,

\begin{abstract}
The solvation of carbon monoxide (CO) in liquid water is important for understanding its toxicological effects and biochemical roles. In this paper, we use ab initio molecular dynamics (AIMD) and CCSD(T)-F12 calculations to assess the accuracy of the Straub and Karplus molecular mechanical (MM) model for CO(aq). The CCSD(T)-F12 CO--\ce{H2O} potential energy surfaces show that the most stable  structure corresponds to water donating a hydrogen bond to the C center. The MM-calculated surface it incorrectly predicts  that the O atom is a stronger hydrogen bond acceptor than the C atom. The AIMD simulations indicate that CO is solvated like a hydrophobic solute, with very limited hydrogen bonding with water. The MM model tends to overestimate the degree of hydrogen bonding and overestimates the atomic radius of the C atom. The calculated Gibbs energy of hydration is in good agreement with experiment (9.3 \kj\ calc. vs 10.7 \kj\ exptl.). The calculated diffusivity of \ce{CO}(aq) in TIP3P-model water was $5.19\times 10^{-5}$ cm$^2$/s calc., more than double the experimental value of $2.32\times 10^{-5}$ cm$^2$/s.
\end{abstract}

\pacs{61.20.Ja}% PACS, the Physics and Astronomy
\maketitle

%\tableofcontents

\section{\label{sec:intro}Introduction}

Carbon monoxide (\ce{CO}) is a highly toxic gas,\cite{Romao2012} with a OSHA Permissible Exposure Limit of only 50 ppm.\cite{osha_co} This toxicity originates from the coordination of \ce{CO} to metalloproteins like hemoglobin.\cite{Prockop2007122} Over the last two decades, \ce{CO} has also been identified as a gasotransmitter that serves as an endogenous signaling molecule in trace concentrations.\cite{Mustafa2009,Motterlini2010} This has spurred the development of \ce{CO}-releasing molecules (CORMs) to allow controlled delivery of carbon monoxide to cellular targets.\cite{Motterlini1142,Schatzschneider2011,Schatzschneider2015,Simpson2016}

The intermolecular interactions of \ce{CO} are remarkably complex. \ce{CO} possesses a modest dipole moment ($\mu_0 = 0.12$ D),\cite{Muenter1975} but has a large negative quadrupole moment ($\Theta_{zz}= (-8.77 \pm 0.31) \times 10^{-40}$ C m$^2$).\cite{Chetty2011} The origin of this quadrupole moment is apparent in the calculated electrostatic potential (ESP) of CO; the ends of the molecule have a negative ESP due to the  C and O lone pairs, while the molecular surface in the vicinity of the C atom has a positive ESP.\cite{Kim2015}

Straub and Karplus developed an early molecular mechanical (MM) model of \ce{CO} for molecular dynamics (MD) simulations of \ce{CO}--myoglobin dissociation.\cite{Straub1991} The model has a Lennard-Jones terms centered at the C and O atoms. Electrostatic interactions in this model are described using 3 point charges; charges are placed on both the C and O atoms as well as one on the C--O bond midpoint. This model has been used extensively to model \ce{CO} dynamics in biomolecular systems since then,\cite{Zheng1996,Meller1998789,Vitkup1997,Lambry1999,Sagnella1999,Sagnella1999,Bossa20043855,Abramo2009,Brinkmann06092016} although it has not been validated by modern methods.

The structure and energetics of solute--solvent interactions can be challenging to study experimentally, so it can be difficult to validate empirical models like this based on experimental quantities alone. Quantum chemical methods provide another approach for  validating molecular mechanical models. High-level ab initio methods like CCSD(T) can now be used routinely to calculate accurate interaction energies of small molecules. Similarly, ab initio molecular dynamics (AIMD) can be used to generate a first-principles comparison the structure liquids and solutions by calculating the energy and forces using a quantum chemical method (e.g., density functional theory, DFT). These methods are valuable for providing first-principles estimates of the solvation structure and can be used to assess aspects of MM models.\cite{Whitfield2007,FIRES,MgZn2013,Riahi2013, Riahi2014a,Adluri2015} 

In this paper, we compare the solvation of \ce{CO} in liquid water calculated using the molecular mechanical models of Straub and Karplus and an ab initio molecular dynamics simulation. In support of this, we also calculate the potential energy surfaces (PES) for a water molecule serving as a hydrogen bond donor and acceptor to a \ce{CO} molecule using CCSD(T)-F12. The solvation energy and diffusivity of the MM model in TIP3P-model water are also compared to experiment.

\section{\label{sec:theory}Theory and Methods}
\subsection{Ab Initio Calculations}
The DFT and CCSD(T)-F12 \ce{CO}--\ce{H2O} potential energy surfaces and interaction energies were calculated using TURBOMOLE v7.0 \cite{TURBOMOLE70}. The CCSD(T) surface used the explicitly correlated F12 method \cite{Hattig2010} using a density-fitting basis set. The aug-cc-pVQZ basis set was used for all atoms.\cite{cc-pVTZ-1st,aug-cc} The DFT calculations were performed using the PBE exchange-correlation functional\cite{PBE} and the aug-cc-pVTZ basis set. The Grimme D3 correction for dispersion was applied.\cite{Grimme2010} The TURBOMOLE input file is included in Supplementary Information.

\subsection{Molecular Mechanical Calculations}

\subsubsection{Parameters and Simulation Cell}

Calculations of the MM \ce{CO}--\ce{H2O} potential energy surfaces and the MD simulations structure of \ce{CO}(aq) were performed using CHARMM c40b2.\cite{CHARMM2009} The parameters for the Straub and Karplus model for \ce{CO} are given in Table \ref{tab:mm_param}. The water molecules were represented using the CHARMM variant of the TIP3P-model.\cite{Jorgensen1983,m_tip3p}  

For the simulations of \ce{CO} in bulk water, a cubic simulation cell was used with an average cell length of 30.8 \AA. Lennard-Jones interactions were scaled to zero using a switching function over the 10--12 \AA\ range. Electrostatic interactions were calculated using the Particle Mesh Ewald (PME) method using a $32 \times 32 \times 32$ grid.\cite{PME} The simulation cell contained 995 water molecules and 1 \ce{CO} molecule.  The simulations to calculate the solvation energies and the radial distribution functions (rdf) used a Langevin thermostat ($\gamma = 5$ ps$^{-1}$) and an Andersen--Hoover barostat.\cite{AndersenBarostat,HooverBarostat} The CHARMM input file is included in Supplementary Information.

\subsubsection{Gibbs Energy of Hydration Calculations}

The Gibbs energy of hydration for \ce{CO} was calculated using the staged decoupling protocol of Deng et al.\cite{Deng2004,DengReview2009}. The electrostatic and dispersion components of the hydration energy were calculated by an 11-window thermodynamic integration simulation ( $\lambda=0.0,0.1,0.2,0.3,0.4,0.5,0.6,0.7,0.8,0.9,$ and $1.0$).

The two states, denoted A and B, correspond to the states where the given solvent--solute interaction is calculated or neglected, respectively. These states are coupled by the thermodynamic integration (TI) variable $\lambda$, through the definition of a linearly-interpolated potential,

\begin{equation}
\mathcal{V}(\lambda) =(1-\lambda)\mathcal{V}_A + \lambda \mathcal{V}_B , 
\end{equation}
Here $\mathcal{V}_A$ and $\mathcal{V}_B$ are the potential energies of the A and B states. 

The repulsive component of the Gibbs energy was calculated using a staged procedure where the repulsive component of the solvent--solute Lennard-Jones interaction potential was reduced to zero using a 9 stage free energy perturbation (FEP) approach.\cite{Deng2004,DengReview2009}

For each TI or FEP simulation, the system was equilibrated for 1 ns followed by a 2 ns sampling simulation. The Gibbs energies were calculated using the Weighted Histogram Analysis Method (WHAM).\cite{Kumar-WHAM}  The reported Gibbs energy was calculated from an average of three independent simulations. 

\subsubsection{Diffusion Calculations}
Simulations used to calculate the \ce{CO} diffusion coefficient were performed with a Nos{\'e} thermostat with a response time of 0.1 ps. The systems were equilibrated for 1 ns before a 2 ns trajectory was collected. The reported diffusivity was calculated  from an average of three independent simulations. The \ce{CO} diffusion coefficient ($D_\textrm{PBC}$) under periodic boundary conditions was calculated using the Einstein relation,

\begin{equation}
D_\textrm{PBC}=\frac{1}{6t}\langle |r_{i}(t)-r_{i}(0)|^{2} \rangle,
\end{equation}
where $r(t)$ is the position of the molecule at time, $t$.  

Yeh and Hummer found that the viscosity of a liquid simulated under periodic boundary conditions depends on the size of the system, which spuriously lowers the calculated diffusivities.\cite{Yeh2004} A correction was applied to our calculated diffusivities using,

\begin{equation}
\label{eqn:einstein2}
D = D_\textrm{PBC} + 2.837297 \frac{k_\textrm{B}T}{6 \pi \eta L} .
\end{equation}
Here, $\eta$ is the solvent viscosity and $L$ is the length of the simulation cell.

\begin{table}
\caption{\label{tab:mm_param}Parameters for the Straub and Karplus\cite{Straub1991} molecular mechanical model for \ce{CO}}
\begin{tabular}{cr}
\hline
parameter   & value  \\ \hline
$q_C$ (e) & -0.75 \\
$q_O$ (e) & -0.85  \\
$q_{COM}$ (e) & 1.6  \\
$\epsilon_{CC}$ (\kj) &  0.109647 \\ 
$\epsilon_{OO}$ (\kj)&  0.6658335 \\ 
$\sigma_{CC}$ (\AA) & 3.83 \\ 
$\sigma_{OO}$ (\AA) & 3.12 \\
$k_{C\equiv O}$ (\kj\ \AA$^{-2}$) & 4666 \\
$r_e$ (\AA) & 1.128 \\ \hline 
\end{tabular}
\end{table}

\subsection{AIMD Simulations}
The AIMD simulations of aqueous \ce{CO} were performed using CP2K version 2.6.\cite{cp2k} The MOLOPT-TZVP basis set was used for all atoms.\cite{VandeVondele2007} The PBE exchange-correlation functional was used \cite{PBE} with the Grimme D3 correction for dispersion.\cite{Grimme2010} The simulation was initiated from an equilibrium structure calculated using the MM model. A canonical-ensemble simulation (NVT) was performed using Langevin dynamics with a bath temperature of 298.15 K and a friction coefficient ($\gamma$) of 1 ps$^{-1}$. A 1 fs time step was used. The O--H bonds were constrained to 0.96 \AA\ using the SHAKE algorithm.\cite{SHAKE} The cell contained one \ce{CO} molecule and 190 water molecules. The cell dimensions were 17.8 \AA\ $\times$ 17.8 \AA\ $\times$ 17.8 \AA. A 15 ps equilibration simulation was performed prior to a 50 ps simulation. The 50 ps simulation was used to calculate the RDF. The CP2K input file is included in Supplementary Information.

\section{\label{sec:results}Results and Discussion}
\subsection{\label{subsec:electric}Electric Properties}

\begingroup
\squeezetable
\begin{table}
\caption{\label{tbl:electric}The calculated and experimental electric properties of \ce{CO}. The MM model is the 3 point model of Straub and Karplus. The DFT results were calculated using the PBE XC functional and the aug-cc-pVTZ basis set. The CCSD(T) calculations were performed with the aug-cc-pVQZ basis set.}
\begin{tabular}{lcc}
\hline
method & $\mu_0$ & $\Theta_{zz}$ \\ 
         &    (D)  & ($\times 10^{-40}$ C m$^2$)\\ \hline
MM       & 0.27\footnote{The negative pole of the dipole of the MM model is on the O end of the CO, opposite to the QM models and experimental data.} & $-8.17$ \\
DFT      & 0.19   & $-6.72$ \\
CCSD(T)  & 0.12   & $-6.58$ \\
exptl.   & 0.12   & $-6.47 \pm 0.13$ \\ 
\hline
\end{tabular}
\end{table}
\endgroup

The models predict significantly different values for the electric moments of CO (Table ~\ref{tbl:electric}). The CCSD(T) results are in the closest agreement with experiment; the dipole moment is equal to the experimental value to two decimal places ($\mu_0=0.12$ D) and the quadrupole moment is predicted correctly within the uncertainty of the experimental value. The DFT model overestimates both moments, predicting a dipole moment of 0.19 D (a 58\% overestimate) and a quadrupole of $-6.72 \times 10^{-40}$ C m$^2$ (a 3 \% overestimate). This performance is typical for the PBE functional.\cite{Hickey2014} Both QM methods predict the correct direction of the dipole vector, where the positive pole is on the O end of the molecule and the negative pole is on the C end (i.e., $^-$CO$^+$).\cite{Meerts1977}

The electric moments of the MM model are in poorer agreement with experiment. The magnitude of the dipole is significantly overestimated ($\mu_0=0.27$ D) and its direction points from the O atom to the C atom, ($^+$CO$^-$). The quadrupole moment of this model is $-8.17 \times 10^{-40}$ C m$^2$ , 26\% larger than the experimental value.

\begin{figure}
\centering
\includegraphics[width=3.25in]{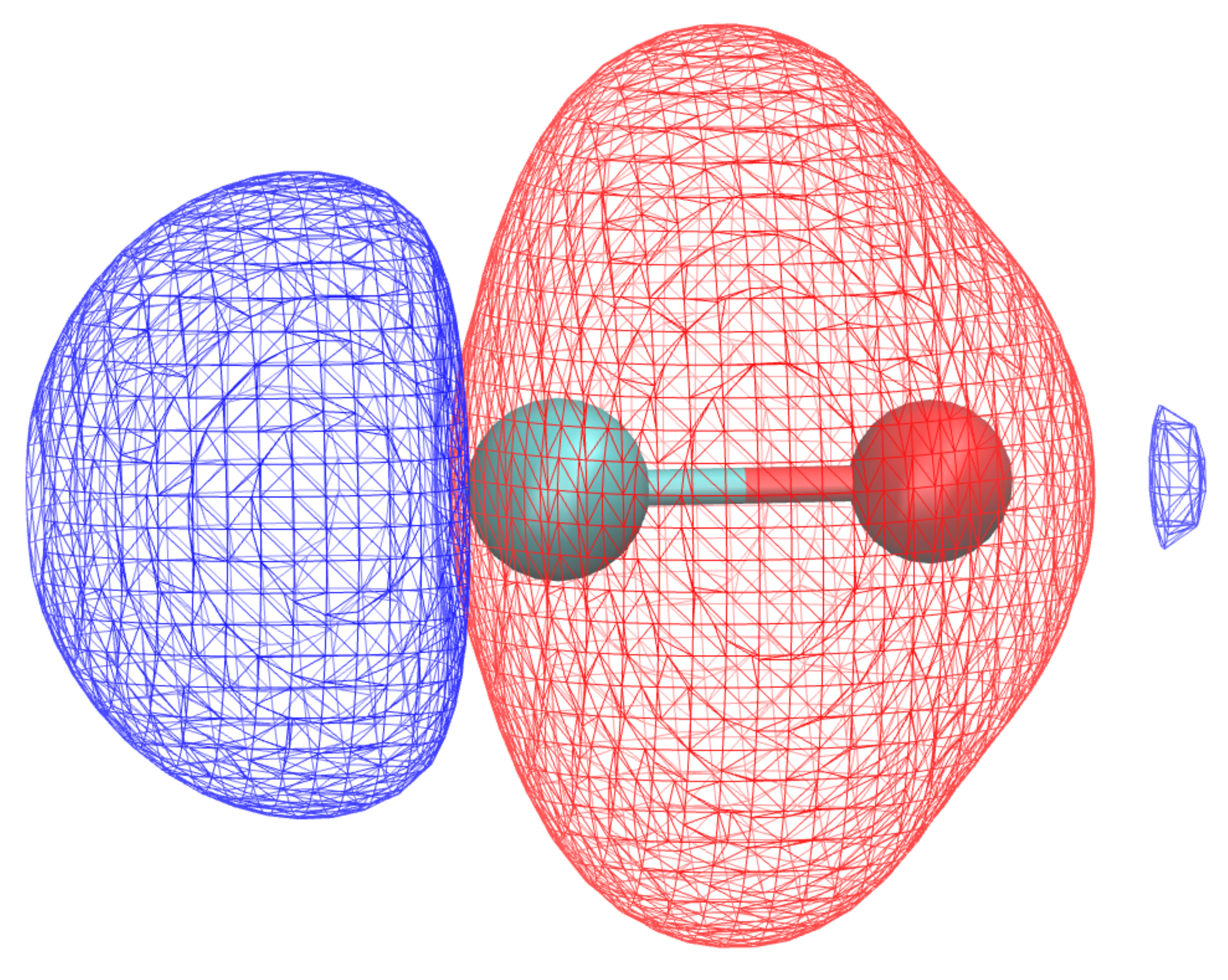}
\caption{The electrostatic potential surface of \ce{CO} calculated using MP2/aug-cc-pVTZ. The blue isosurface corresponds to a negative ESP that interacts favorably with positive charges ($ESP(r)=-0.006$ a.u.). The red isosurface corresponds to a positive ESP that interacts favorably with negative charges ($ESP(r)=+0.006$ a.u.).}
\label{fig:ESP}
\end{figure}

The off-center charge of the MM model is located at the bond midpoint.  Although this simplifies the implementation and parameterization, this does not reflect the true electronic distribution in \ce{CO}. The electrostatic potential is positive in the space near the nuclei and negative in the space on opposite ends of the molecule. The negative ESP lobe adjacent to the C atom is particularly large. This is consistent with the MO theory description of \ce{CO}, where the HOMO is a $\sigma$ bond with a large lobe protruding from the C atom along the bond axis.\cite{Bochmann1994} Based on this, a more realistic representation of the ESP would be achieved if the off-center charge was negative and located in the space opposite to the C atom.\cite{Kim2015} This type of off-center charge has been successfully used to describe the electrostatic interactions of molecules with a $\sigma$-hole,\cite{halogenbond_ff1,halogen_bonds_ff,halogen_bond_opls,Adluri2015} although not all molecular simulation codes support this type of site at present.

\subsection{\label{subsec:optstruct}Optimized Structures}

\begin{figure}
\centering
\includegraphics[width=3.25in]{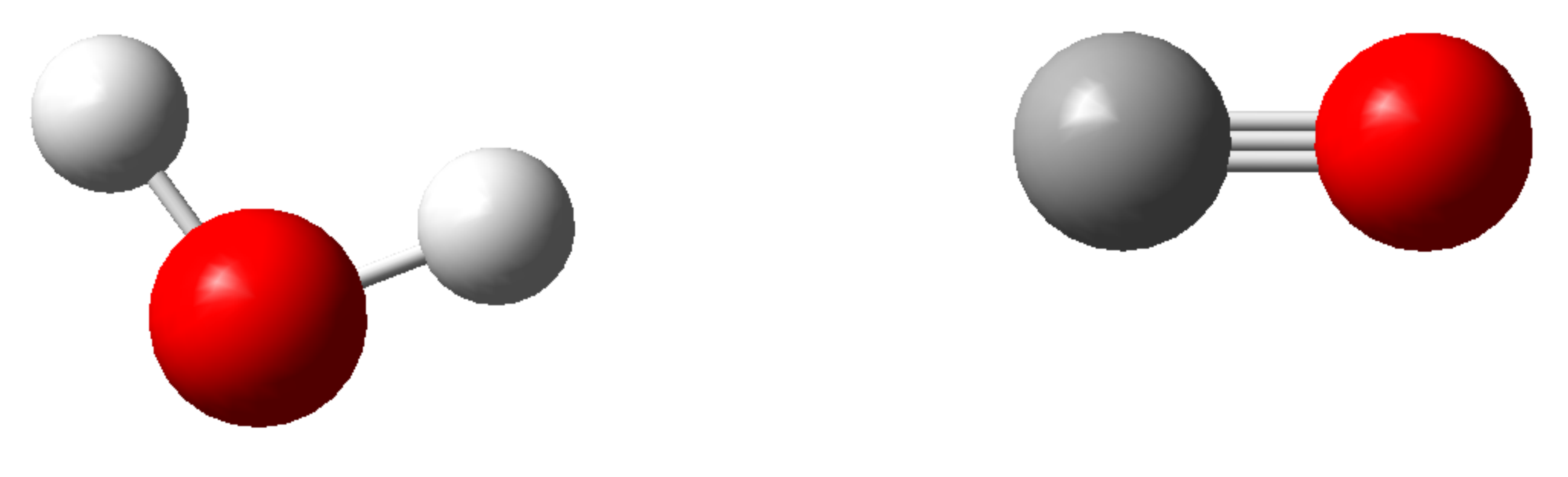}
\caption{\label{fig:co_H2O}Minimum energy structure of the CO--\ce{H2O} complex. The most stable structure for the CCSD(T) and DFT structures corresponds to the water molecule donating a hydrogen bond to the C atom of the monoxide.}
\end{figure}

\begingroup
\squeezetable
\begin{table}
\caption{\label{tbl:interaction}The calculated CO--\ce{H2O} interaction energies using the MM model of Straub and Karplus, the DFT model (PBE/aug-cc-pVTZ), and  CCSD(T) (CCSD(T)-F12/aug-cc-pVQZ//MP2/aug-cc-pVTZ).}
\begin{tabular}{l c}
\hline
method & $\Delta E$ (kJ/mol) \\  \hline
MM           & $-5.1$ \\
DFT          & $-8.9$ \\
CCSD(T)-F12  & $-9.6$ \\ \hline
\end{tabular}
\end{table}
\endgroup

The minimum energy structure of the \ce{H2O}--\ce{CO} interactions corresponds to the C atom of the \ce{CO} serving a hydrogen bond acceptor (Figure ~\ref{fig:co_H2O}). The interaction energies calculated using the three models are collected in Table ~\ref{tbl:interaction}. The CCSD(T)-F12 predicts a modest interaction energy of $-9.6$ \kj. The DFT-calculated interaction energy is slightly weaker, with an interaction energy of $-8.9$ \kj. The interaction energy calculated using the MM model is only $-5.1$ \kj, significantly weaker than the CCSD(T)-F12 interaction energy. The reversed polarity of the dipole in the MM model causes this minimum to be higher energy hydrogen bond to the O atom with an interaction energy of $-10.9$ \kj. 

\subsection{\label{subsec:pes}Potential Energy Surfaces}
Two-dimensional potential energy surfaces were calculated for the interaction between CO and \ce{H2O} molecules. The CO and \ce{H2O} molecules were fixed at their experimental gas phase structures. The coordinates were defined in terms of the distance between the center of the C--O bond and the O--C --- \ce{O(H2)} angle. In the first surface, the vector bisecting the  H--O--H angle points towards the midpoint of \ce{CO}, with the water O atom oriented towards the \ce{CO} (donor).  In the second surface, one of the O--H bonds of the water molecule points towards the center of the \ce{CO} molecule (acceptor). 

\begin{figure}[h]
\centering
\includegraphics[width=3.25in]{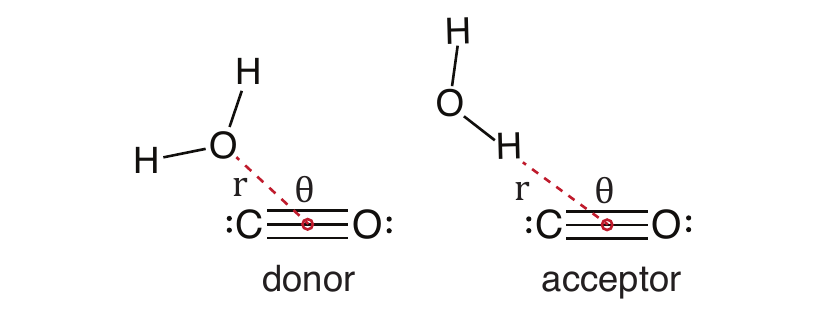}
\label{fig:bond-scheme}
\end{figure}

\begin{figure}
\centering
\includegraphics[width=3.25in]{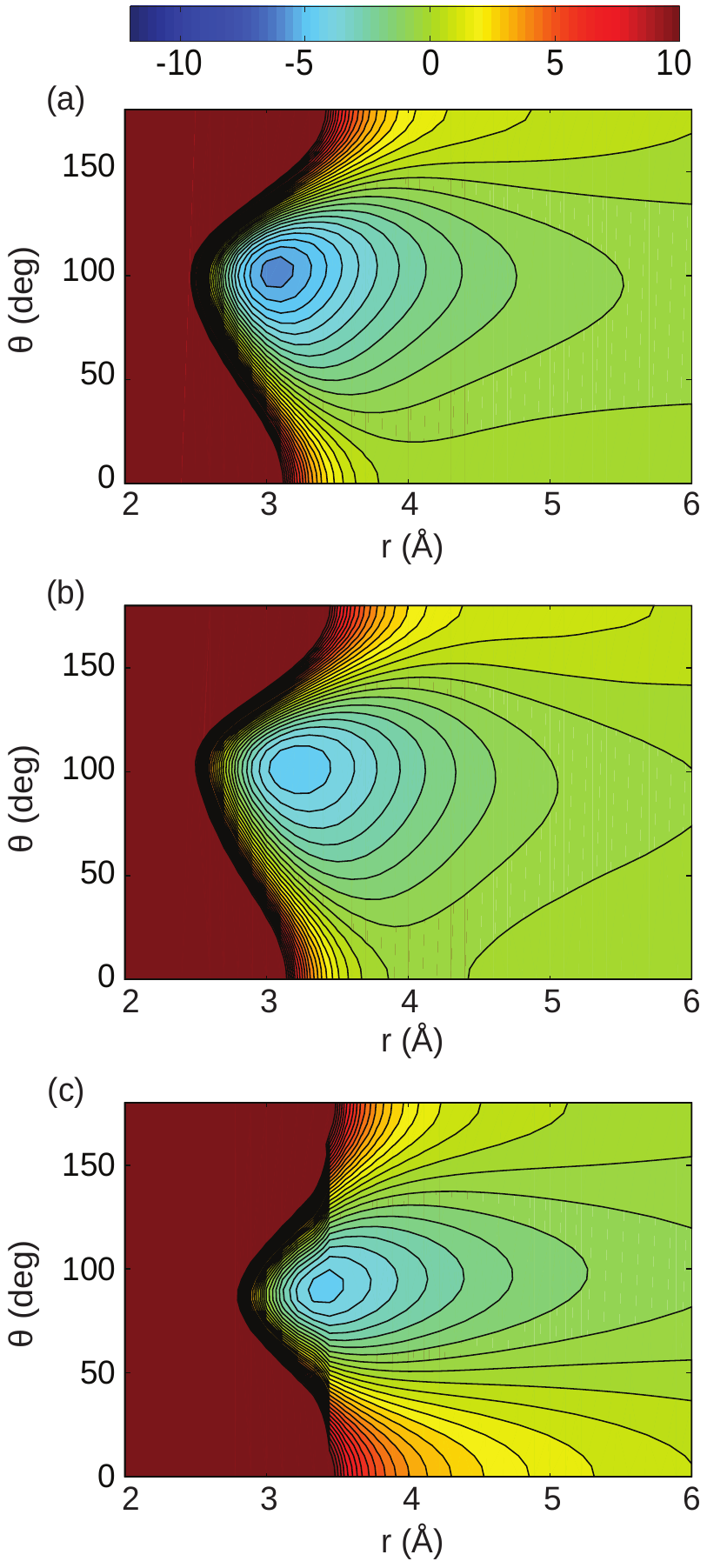}
\caption{Potential energy surfaces of the CO--\ce{H2O} interaction where the bisector of the H--O--H angle is directed towards the center of geometry of CO. The surfaces are calculated using (a) CCSD(T)-F12/aug-cc-pVQZ, (b) PBE-D3/aug-cc-pVTZ, and (c) Straub and Karplus MM model.}
\label{fig:donor-pes}
\end{figure}

\begin{figure}
\centering
\includegraphics[width=3.25in]{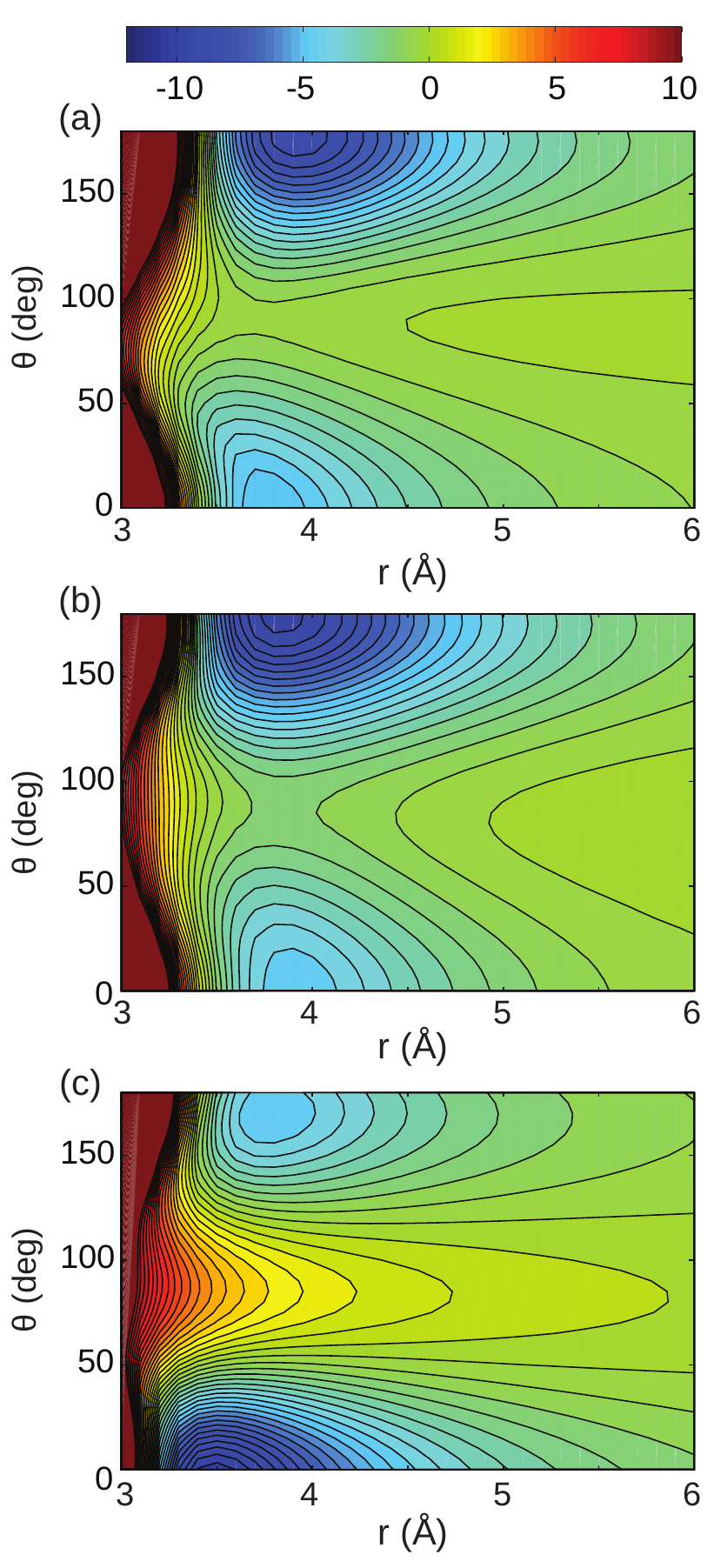}
\caption{Potential energy surfaces of the CO--\ce{H2O} interaction where one of the O--H bonds is directed towards the center of geometry of CO. The surfaces are calculated using (a) CCSD(T)-F12/aug-cc-pVQZ, (b) DFT (PBE-D3/aug-cc-pVTZ), and (c) Straub and Karplus MM model.}
\label{fig:acceptor-pes}
\end{figure}

The calculated potential energy surfaces for both donor and acceptor surfaces using the CCSD(T)-F12, DFT, and MM models are represented in Figs. \ref{fig:donor-pes} and \ref{fig:acceptor-pes}. For consistency and to reduce the surface to 2D, both the \ce{H2O} and the \ce{CO} molecules were maintained at their rigid experimental gas phase geometry in all configurations. The DFT PES provides an indication of the accuracy of the AIMD simulations because the same exchange-correlation functional (PBE) is used to calculate the potential energy of the CO--water simulation cell in the AIMD simulations and also use a triple-$\zeta$ basis set.

The qualitative features of the surfaces calculated with the three models are generally consistent. For the calculated donor surface plots (Figure ~\ref{fig:donor-pes}), the region of lowest energy minimum corresponds to a dipole--quadrupole interaction between the \ce{H2O} and the monoxide, where the interaction is strongest at the midpoint of the \ce{CO} bond. The location of the minimum is approximately $\theta=100$\degree\ and $r=3.2$ \AA\ in both QM surfaces. The stability of this mode of interaction is modest; on the CCSD(T)-F12 surface, the minimum is roughly $-6$ \kj. This minimum is slightly higher in energy in the DFT and MM models, with energies in the $-4$ -- $-5$ \kj\ range.  The location of the MM minimum is somewhat different than the QM surfaces; it is centered at $\theta=90$ \degree\ and $r=3.4$ \AA.

There are two minima on the acceptor surfaces. The minimum at $\theta=180$ \degree\ corresponds to the C atom of the monoxide molecule accepting a hydrogen bond from the water along the C--O bond axis, while the minimum at $\theta=0$\degree\ corresponds to the O atom of the monoxide molecule accepting a hydrogen bond from the water along the C--O bond axis. In the CCSD(T)-F12 and DFT surfaces, the minimum around $\theta=0$ \degree\ is higher energy than the $\theta=180$ \degree\ minimum; the potential energy of about $E_\textrm{min}=-5$ \kj. 

The most significant difference of the acceptor surface calculated using the MM model in comparison to the CCSD(T)-F12 is that the minimum corresponding to donation to the C atom is less stable ($-5$ \kj) and the minimum corresponding to donation to the O atom is more stable ($-10$ \kj) surface. The reversal of the stability of these minima is a consequence of the incorrect polarity of the dipole of the MM model, where the negative end is on the O atom and the positive end is on the C atom.

\subsection{\label{subsec:rdf}Radial Distribution Functions}

\begin{figure}
\centering
\includegraphics[width=6.5in]{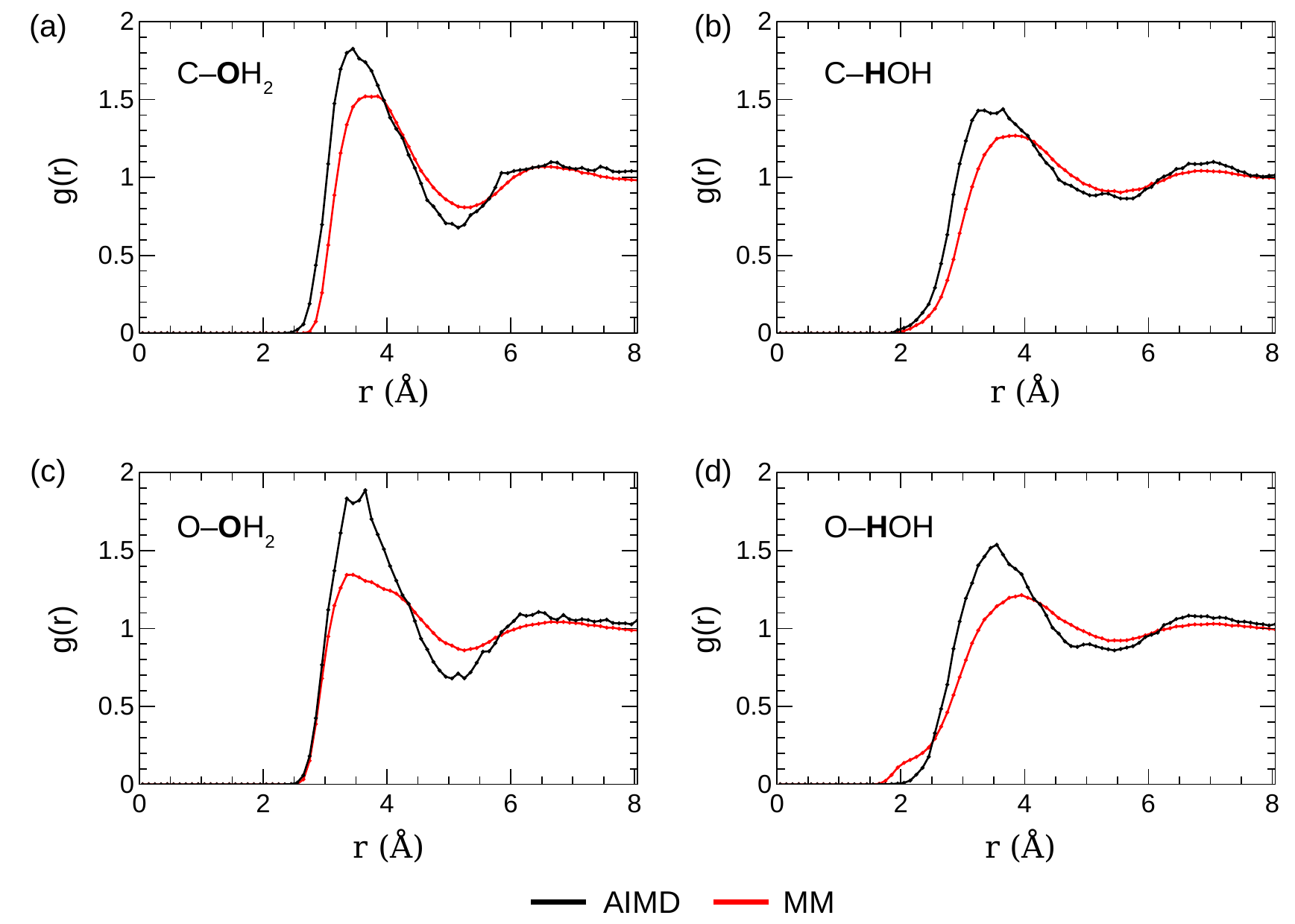}
\caption{Radial distribution functions of CO(aq) calculated from AIMD and MM trajectories. (a) C--\ce{O(H2)}, (b) O--\ce{O(H2)}, (c) C--HOH, and (d) O--HOH.}
\label{fig:rdf}
\end{figure}

To examine CO--\ce{H2O} interactions in bulk water, radial distribution functions were calculated from the AIMD and MM trajectories of \ce{CO}(aq). Figure \ref{fig:rdf} (a) shows the rdf between the water O atom and the monoxide C atom and Figure \ref{fig:rdf} (c) shows the rdf between the water O atom and the monoxide O atom. Figure \ref{fig:rdf} (b) and (d) shows the rdfs of C and O ends with water hydrogen atoms, which can be used to infer the formation of \ce{CO}--\ce{H2O} hydrogen bonds.

The most significant difference between these two models is that the first peak of the  C--\ce{O(H2)} rdf occurs at a 0.1 \AA\ larger distance than the AIMD value. This suggests that the Lennard-Jones radius of the monoxide carbon atom in the MM model is too large. This is consistent with the potential energy surfaces plotted in Figure \ref{fig:donor-pes}, where the MM surface is more strongly repulsive at C--\ce{O(H2)} distances in the 3--4 \AA\ range in comparison to the CCSD(T)-F12 surface. 

The position of the first peak of the MM O--\ce{OH2} rdf agrees well with the AIMD result, suggesting that the MM Lennard-Jones radius of O is reasonable. The first coordination sphere of the rdfs calculated from the AIMD simulations are more pronounced than in the MM simulations, although this aspect of AIMD simulations of liquid water is not entirely reliable due to a trend of over-structuring liquid water at STP.\cite{Parrinello2009,Rothlisberger2012,Schwegler2016} 

\begin{figure}
\centering
\includegraphics[width=3.25in]{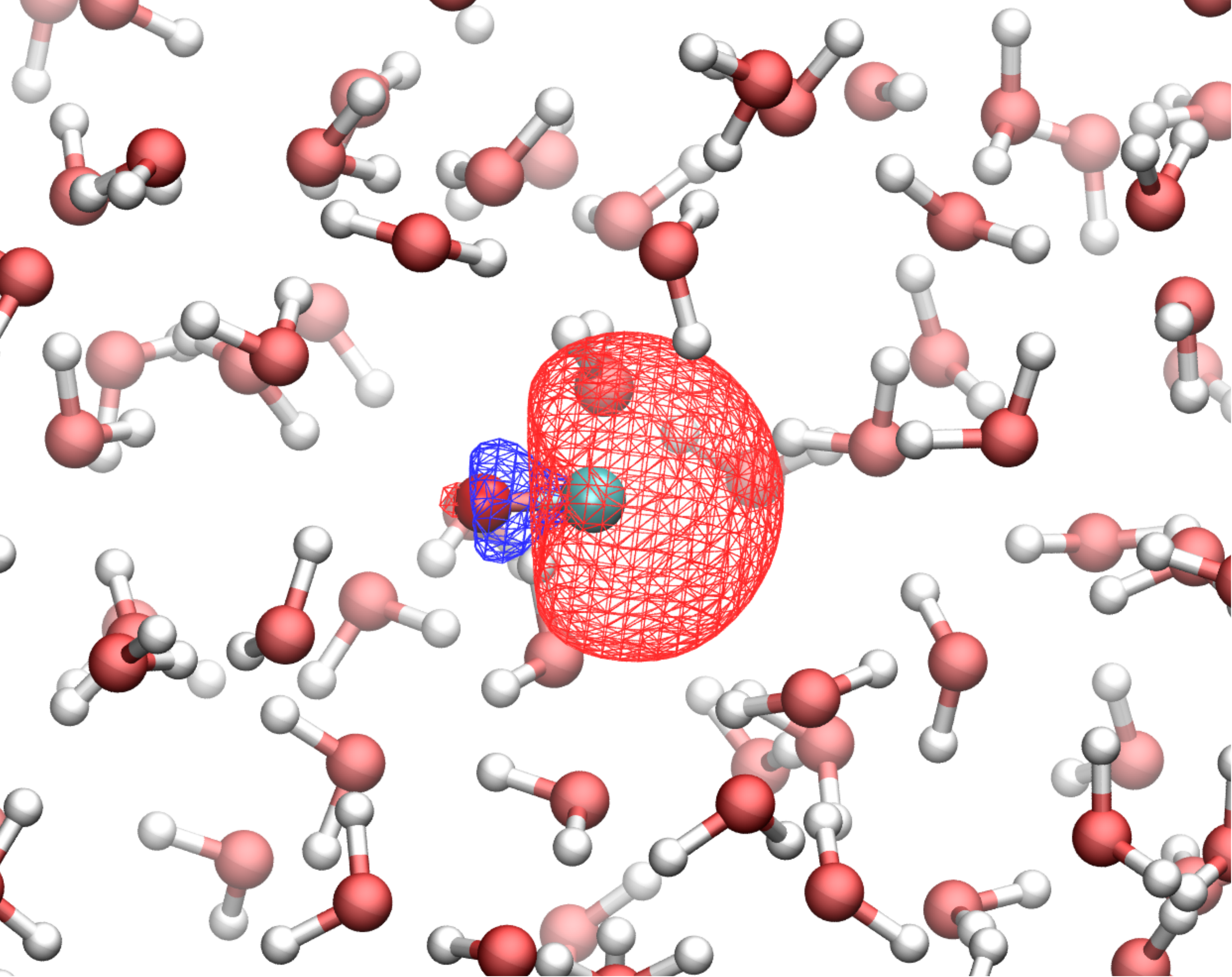}
\caption{The Wannier-localized orbital of the AIMD simulation of \ce{CO}(aq) corresponding to the carbon-centered lone pair. The MO is rendered by the red and blue mesh surfaces. The surfaces are plotted at an isocontour value of $| \psi | = 0.016$ a.u.}
\label{fig:MO}
\end{figure}

Both models have very modest shoulders on the left-side of the first peak of the rdf for the distance between the monoxide C-atom and the water hydrogen  (Figure \ref{fig:rdf} (c)). This indicates that neither model predicts that there are strong hydrogen bonds between the monoxide carbon and water molecules in aqueous solutions. This is apparent in the Wannier-localized orbitals,\cite{Ambrosetti2016} which shows carbon-centered lone pair to be too distant from the water hydrogen atoms to serve as a strong hydrogen bond acceptor (Figure ~\ref{fig:MO}).

The first peak of the MM RDF for the monoxide oxygen and water hydrogen distance (Figure \ref{fig:rdf} (d)) has a shoulder on the left side between 1.9 and 2.1 \AA\ that is consistent with a water molecule donating a hydrogen bond to the monoxide oxygen. This shoulder is much more pronounced in the rdf calculated from the MM simulation and is a minor feature in the AIMD rdf. This is consistent with the QM potential energy surfaces, which show that the MM model overestimates the strength of hydrogen bonds where the monoxide oxygen is the acceptor. 

\subsection{Hydration Energy}

\begingroup
\squeezetable
\begin{table}
\label{tbl:fep}
\caption{The Gibbs energy of hydration of the MM model for CO. All values are in \kj. The experimental value is estimated from the solubility constant reported in Ref. \citenum{CO_solubility}. The standard state used is 1 M gas $\rightarrow$ 1 M solution.}
\begin{tabular}{l c}
\hline
component & $\Delta$G (kJ/mol) \\ \hline
electrostatic & $-2.8$ \\
dispersion   &  $-15.5$ \\
repulsive  & $29.1$ \\ \hline
total & $10.7 \pm 0.2$ \\ \hline
exptl. & $9.3$ \\
\hline
\end{tabular}
\end{table}
\endgroup

To quantify how accurately this model predicts the CO--water interactions, the Gibbs energy of hydration of CO was calculated using alchemical FEP (Table ~\ref{tbl:fep}). This analysis indicates that the hydration energy calculated using the MM model is in reasonably good agreement with experiment ($10.7$ \kj\ vs $9.3$ \kj, respectively). The electrostatic interactions between CO and water are weak, contributing only $-2.8$ \kj\ to the hydration energy. The component from dispersion is considerably stronger, contributing $-15.5$ \kj. These interactions are counteracted by a large repulsive energy of 29.1 \kj. These interactions are generally consistent with the hydration of a non-polar solute.\cite{Riahi2014a}

\subsection{\label{sec:diffusion}Diffusion Coefficient}

The calculated diffusivity of \ce{CO} in liquid water at 30 ~\degree C was calculated using the finite-size corrected Einstein relation (Eqn ~\ref{eqn:einstein2}. A diffusivity of  $D=5.19 \pm 0.82 \times 10^{-5}$ cm$^2$/s was calculated. This is considerably larger than the experimental value of $D=2.32 \pm 0.07 \times 10^{-5}$ cm$^2$/s.\cite{ExptlDiffusion}. The TIP3P water model predicts a viscosity that is significantly lower than the experimental value, so by the Stokes--Einstein equation, the diffusivity of the solutes is this medium will be overestimated.\cite{ViscosityTIP3P} A \ce{CO} model developed for use with water models like TIP4P-2005\cite{Tazi2012} or TIP4P-FB\cite{ForceBalance} that have more accurate viscosities could provide improved diffusivities.

\section{\label{sec:conclusions}Conclusions}
AIMD simulations, CCSD(T)-F12 calculations, and MM MD simulations were used to study the solvation of \ce{CO} in liquid water. The AIMD simulations indicated that there are no persistent water--CO hydrogen bonds, although the CCSD(T) potential energy surfaces and AIMD rdfs indicated that \ce{CO} forms its strongest interactions with water when the lone pair on the C atom serves as a hydrogen bond acceptor. The weakness of these interactions is consistent with the poor solubility of \ce{CO} in liquid water.

These QM calculations were compared to the results from the MM model of Straub and Karplus. This model overestimates the dipole and quadrupole moments and predicts the opposite polarity of the dipole moment. This causes hydrogen bonds to the monoxide O atom to be the preferred interaction, while the QM calculations indicate that hydrogen bonds to the C atom are stronger. In comparison to the AIMD results, the atomic radius of the MM monoxide C atom model appears to be slightly too large by roughly 0.05 \AA. Despite these limitations, the \dGhydr\ of this model is in very good agreement with the experimental value (9 \kj\ calc. vs 10 \kj\ exptl.). The diffusivity the MM model is significantly higher than the experimental value ($5.19\times 10^{-5}$ cm$^2$/s calc. vs $2.32\times 10^{-5}$ cm$^2$/s exptl.). 

An improved MM model could likely be designed by repositioning the off-center charge to be located in the place of the C-centered lone pair and changing the charges to reflect the actual electric moments. The Lennard-Jones radius of the C atom should also be increased. Improving the diffusivity would require developing a model that can be used with a water model with an accurate viscosity.

\section*{\label{sec:si}Supplementary Material}
See supplementary material for CP2K, CHARMM, and TURBOMOLE input files.

\section*{\label{sec:acknowledgements}Acknowledgements}
The authors thank NSERC of Canada for funding through the Discovery Grant program (Application 418505-2012). EAW thanks the School of Graduate studies at Memorial University for a graduate fellowship. EAW also thanks ACENET for an Advanced Research Computing Fellowship. Computational resources were provided by Compute Canada (RAPI: djk-615-ab) through the Calcul Quebec and ACENET consortia.

%\section*{\label{sec:references}References}

\bibliographystyle{aipnum4-1}
\bibliography{ref}

\end{document}